# Analytical Model for the Transient Permittivity of Uncured TiO$_2$ Whisker/Liquid Silicone Rubber Composites Under an AC Electric Field

Zikui Shen, Zhenyu Xin, Xilin Wang, Xinyu Wei, Zhidong Jia[*]

Engineering Laboratory of Power Equipment Reliability in Complicated Coastal Environment, Shenzhen International Graduate School, Tsinghua University. Shenzhen 518055, P.R. China
[*]jiazd@sz.tsinghua.edu.cn

**Abstract:** The electric field grading of dielectric permittivity gradient devices is an effective way of enhancing their insulation properties. The in-situ electric field-driven assembly is an advanced method for the fabrication of insulation devices with adaptive permittivity gradients, however, there is no theoretical guidance for design. In this work, an analytical model with a time constant is developed to determine the transient permittivity of uncured composites under an applied AC electric field. This model is based on optical image and dielectric permittivity monitoring, which avoids the direct processing of complex electrodynamics. For a composite with given components, the increased filler content and electric field strength can accelerate the transient process. Compared with the finite element method (FEM) based on differential equations, this statistical model is simple but efficient, and can be applied to any low-viscosity uncured composites, which may contain multiple fillers. More importantly, when a voltage is applied to an uncured composite insulating device, the proposed model can be used to analyse the spatiotemporal permittivity characteristics of this device and optimise its permittivity gradient for electric field grading.

## 1. Introduction

In recent years, the development of electrical equipment and power electronic devices with high voltages and large capacities, such as 1000-kV AC gas insulated switchgears (GIS) and transmission lines and 6.5-kV insulated gate bipolar transistors, has occurred. These devices require electrical insulation materials with superior characteristics to increase their operating voltages and decrease space. However, unexpected local increases in the electric field easily trigger partial discharge, electrical treeing, and surface flashover, leading to insulation failure. This problem significantly limits equipment compactness and reduces operation reliability. Meanwhile, graded permittivity materials are effectively used for electric field grading[1]. Several research groups have demonstrated that a graded permittivity can be achieved by lamination[2, 3], centrifugation[4, 5], electrophoretic deposition[6], electric field assembly, magnetophoretic methods[7, 8], and three-dimensional printing[9]. Compared with other techniques, the in-situ electric field assembly represents the most versatile method that is applicable to any insulation system[10]. During the AC electric field grading, the higher the field strength, the greater the required permittivity. During in-situ electric field assembly, the higher the field strength, the stronger are the resulting material property enhancements, including the increase in dielectric permittivity. Hence, these two processes are perfectly correlated.

Electric field assembly is employed to manipulate the motions of colloidal micro- and nanoparticles with a minimum energy configuration. Over the last few decades, this technology has been gradually improved and applied to many fields, including cell separation[11, 12], composite material enhancement[13, 14], and microfluidic chip manufacture[15–17]. It is also used for aligning zero-dimensional particles into one-dimensional chains or orienting one-dimensional particles to enhance the properties of composite materials, such as tensile characteristics, permittivity, and thermal conductivity. Inspired by the performance enhancement of various composites (including composite films), we have explored in-situ nonuniform electric field-induced gradient permittivity materials for field grading with improved insulation characteristics[10]. Unlike the material property enhancement by a uniform electric field, dielectric permittivity is a function of both time and space during the in-situ electric field-assisted preparation. Accurate control of the process parameters is required to obtain a desired permittivity gradient. These parameters (including the permittivity of the filler, voltage, and application time) are strongly coupled to each other and related to the structure shape. Therefore, they are difficult to measure experimentally. Furthermore, no theoretical design framework for such preparation has been developed yet.

To establish such a framework, it is necessary to study the time-domain dynamics of particles in an electric field. While the existing studies successfully simulated the electric field-assisted assembly of particles using the Maxwell stress tensor, this approach is computationally expensive and limited to microscale computational domains. In this work, a new analytical model for the transient permittivity of uncured TiO$_2$ whisker/liquid silicone rubber (LSR) composites exposed to an AC electric field that involves online optical and dielectric permittivity monitoring is proposed. This model can be used to characterise the time-domain variation of the permittivity of uncured composites subjected to an AC electric field and guide the design and fabrication of permittivity gradient devices for electric field grading.

## 2. Theoretical background
### 2.1. Interactions between particles and external electric field

Suspended particles become polarised in an electric field, which results in the separation of positive and negative charges and formation of electric dipoles. The dipole moment of a dielectric sphere can be expressed as[18]

$$p = 4\pi r^3 \varepsilon_0 \varepsilon_m \left(\frac{\varepsilon_p - \varepsilon_m}{\varepsilon_p + 2\varepsilon_m}\right) E \quad (1)$$

where ε0 is the permittivity of free space; $\varepsilon_m$ and $\varepsilon_p$ are the relative permittivities of the suspension medium and particle, respectively; r is the particle radius; and $\frac{\varepsilon_p - \varepsilon_m}{\varepsilon_p + 2\varepsilon_m}$ is the Clausius–Mossotti (CM) factor.

For a rod-like particle, the axial dipole moment is much larger than the dipole moment perpendicular to the axial direction. The former parameter is equal to[19]

$$p_{||} = 2f_D E \cos\theta \quad (2)$$

where $f_D = \frac{\pi r^2 l}{2}\varepsilon_0\varepsilon_m \frac{(\varepsilon_p - \varepsilon_m)^2}{[\varepsilon_m + (\varepsilon_p - \varepsilon_m)L_{||}](\varepsilon_p + \varepsilon_m)}$; r and l are the radius and length of the rod-like particle, respectively; and $L_{||} \approx 4r^2/l^2[\ln(l/r) - 1]$ is the depolarisation factor. The particle axis forms an angle θ with respect to the direction of the applied electric field.

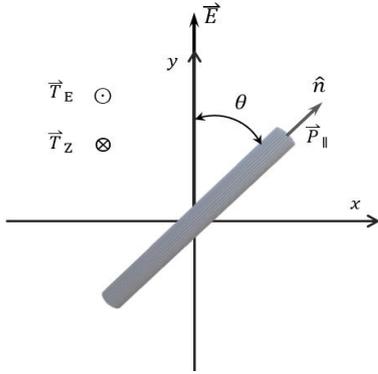

***Fig. 1.*** *Schematic of a rod-like colloidal particle in an applied electric field*

According to the principle of minimum energy, randomly distributed particles move to reduce their potential energy. Thus, rod-like particles tend to become oriented along the electric field, as shown in Fig 1. The time average electro-orientation torque generated by the this field can be expressed as[19]

$$T_E = f_D E^2 \sin 2\theta \quad (3)$$

The equivalent force (force couple) is defined as

$$F_E = \frac{2f_D}{l} E^2 \cos\theta \quad (4)$$

The resultant rotation of the particle depends on the interaction between the electrorotational torque and the drag torque caused by the viscous drag, as shown in Fig. 1. At low Reynolds numbers (<< 1), the viscous term dominates the fluid dynamics. According to Stokes' law, the balanced angular velocity can be written as[20, 21]

$$\omega = \frac{f_D}{f_s} E^2 \sin 2\theta \quad (5)$$

where $f_s = \frac{2\pi\eta(l^2 + 4r^2)l}{3[2\ln(\frac{l}{r}) - 1]}$, and η is the kinematic coefficient of the viscosity.

When the external electric field is non-uniform, dipoles are subjected to a dielectrophoretic force. For a dielectric sphere, its magnitude can be expressed as[18]

$$F_{DEP} = 2\pi r^3 \varepsilon_0 \varepsilon_m \left(\frac{\varepsilon_p - \varepsilon_m}{\varepsilon_p + 2\varepsilon_m}\right)\nabla E^2 \quad (6)$$

For a rod-like particle, (5) should be written as

$$F_{DEP} = 2f_D \cos^2\theta \nabla E^2 \quad (7)$$

These formulas show that the CM factor determines the direction of the dielectrophoretic force. If $\varepsilon_p > \varepsilon_m$, the force is oriented toward the stronger field, and if $\varepsilon_p < \varepsilon_m$, the force is oriented toward the weaker field. Furthermore, $F_{DEP}$ is much smaller than $F_E$ because the electric field in a power system is almost uniform, and its gradient is relatively small.

### 2.2. Electrostatic interactions between particles

In addition, polarised colloidal particles interact with each other, leading to their self-assembly. When the centre line of two particles forms an acute angle with the direction of the electric field, the particles attract each other. In contrast, when the centre line is perpendicular to the electric field, they repel each other, as shown in Fig. 2.

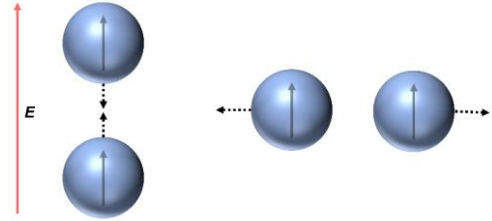

***Fig. 2.*** *Schematic of the electrostatic interactions between two dipoles*

The interactions between the colloidal particles of the same type cause their assembly in chains along the direction of the electric field that minimises the potential energy of the system. The driving force behind this process can be expressed as[22]

$$F_{chain} = -2C\pi r^2 \varepsilon_0 \varepsilon_m (CM)^2 E^2 \quad (8)$$

where the negative sign represents the attractive force, coefficient C depends on the distance between the particles and the chain length, and the value range of C is 3–1000. In particular, if the system contains two or more colloidal particles and their CM coefficients have opposite signs, the particles form alternating chains in the direction of the vertical electric field.

### 3. Experimental

#### 3.1. Materials

Two-component silicone rubber SYLGARD 184 (density: 1.03 g cm$^{-3}$) obtained from Dow Chemical Co. was selected in this work because of its optimal viscosity (3.5 Pa·s at room temperature), curing temperature, and sedimentation time. Moreover, both the rubber and curing agents were optically transparent, which was of primary importance for conducting online optical microscopy observations. Rod-like TiO$_2$ whiskers with a diameter of 300 nm, average length of 5 μm purchased from Haoxinano Co. were used as the filler due to their high dielectric permittivity and aspect ratio. The scanning electron microscope (SEM) image of TiO$_2$ whiskers is displayed in Fig. 3.

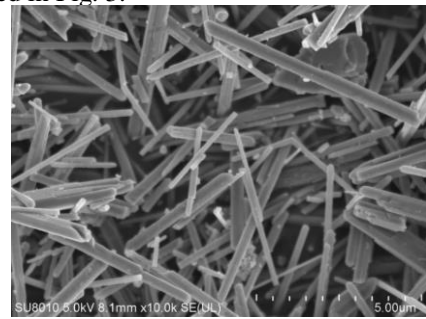

*Fig. 3. An SEM image of the TiO$_2$ whiskers used in this work*

### 3.2. Mixture preparation and measurement procedure

The same experimental procedure was applied for both the optical observation and dielectric characterisation. The A and B rubber components were mixed at a mass ratio of 10:1, after which TiO$_2$ whiskers and coupling agent KH550 (mass ratio of 100:1) were introduced to the resulting mixture by a dispersion machine for 20 min and then degassed for 10 min under vacuum. For optical observations, the mixture was deposited onto a glass plate between two electrodes (Fig. 4a). To conduct dielectric permittivity measurements, the mixture was directly poured into a polytetrafluoroethylene (PTFE) gasket placed between two parallel metal electrode plates, as shown in Fig. 4b.

Particle electrodynamics was studied at a room temperature of approximately 25 °C by optical microscopy using a Phoenix 5000 microscope operated in the transmission mode. A 10-kHz voltage generated by a Suman CTP-2000K power supply was applied to the two parallel electrodes to produce a quasi-uniform electric field of 200 V/mm.

The sample cell used for dielectric permittivity measurements was composed of two flat stainless-steel substrates with a size of 130×130 mm. The inter-electrode gap was controlled by a PTFE circular gasket with an inner diameter of 20 mm, outer diameter of 25 mm, and thickness of 2 mm. A virtual oscilloscope (Hantek 6002BE) was utilised to monitor the U–I signals generated during the voltage application. The equivalent circuit of the measurement device is shown in Fig. 4b. The loop current can be divided into the current passing through the mixture and the current caused by air and edge effects. The former current may be subdivided into resistive and capacitive currents according to the phase difference.

$$I = I_{R1} + I_{C1} + I_{R2} + I_{C2} \tag{9}$$

The relative permittivity of the sample is defined by the formula

$$\varepsilon_s = \frac{I_{C2} h}{\omega \varepsilon_0 U S_2} \tag{10}$$

where h and S$_2$ are the thickness and area of the sample, respectively. In this study, these parameters were measured individually with high accuracy using various techniques, including image recognition. For the correct extraction of dielectric permittivity, it is necessary to obtain the current components of the air and stray capacitance. Hence, we first measured the loop current of the device without a sample under U$_0$ and calculated the effective area of the air as follows:

$$S_0 = \frac{I_{C0} h}{\omega \varepsilon_0 U_0} \tag{11}$$

In the subsequent monitoring process, the dielectric parameters of the air were considered constant. The capacitive current component of the sample is equal to

$$I_{C2} = I_C - \frac{I_{C0}}{U_0} \frac{S_0 - S_2}{S_0} U \tag{12}$$

Finally, we used the described method and a broadband dielectric spectrometer (Novocontrol Concept 40) to measure the permittivity of each cured sample and verified the accuracy of the obtained results.

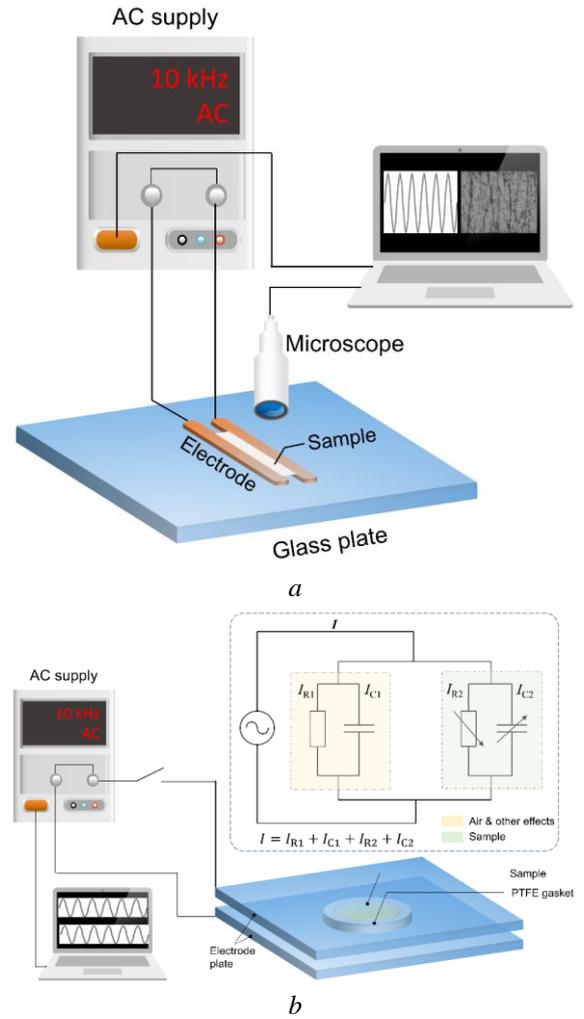

*Fig. 4. Experimental setup.*
*(a) schematic of the online optical observation system,*
*(b) schematic of the online dielectric permittivity monitoring device*

## 4. Results and discussion

### 4.1. Optical monitoring at low filler contents

Optical observations were performed at a low filler ratio of 1 vol.%, and the thickness of the observed mixture was limited to approximately 50 μm by the electrodes located on both sides of the measurement setup to ensure proper light transmission and discrimination.

Fig. 5 shows the distribution of the particles captured at different stages of the voltage application. The particles are randomly distributed across the sample at the initial stage and gradually form local short chains with lengths of approximately 50–150 μm. These short chains attract each other and connect end-to-end to form longer chains, which penetrate the two electrodes (see Figs. 5b and 5c). The resulting long chains are very stable and become further strengthened by adsorbing small particles and short chains located nearby (Figs. 5c and 5d). During this process, the electric field causes particles to orient and form chains, whose degrees of alignment were quantified by Herman's orientation factor (HOF), where 0 and 1 represented the random and perfectly aligned orientations, respectively. All analyses were conducted using the open-source ImageJ software[23–25]. First, a two-dimensional (2D) fast Fourier transformation (FFT) intensity profile (Fig. 6b) was calculated from a greyscale 8-bit optical micrograph cropped to a size of 1024×1024 pixels (Fig. 6a). The obtained FFT intensity profile of the sample image containing strong chains is oblong and anisotropic. It also shows an apparent energy concentration at a specific angle with higher grey and brightness values (Figs. 6c and 6d). In contrast, the FFT intensity profile obtained for the sample image with disordered particles has no directionality, as shown in Figs. 6a and 6b. Additionally, the Oval Profile plug-in was used to sum up the grey values of different radius directions in the range of 0–360° at a given radius, and the obtained distributions of spectral intensities (represented by grey values) along a specified direction are shown in Fig. 7. By analysing the peak position of the spectral intensity distribution, the corresponding orientation angle can be determined with high accuracy. The HOF values were calculated by the formulas

$$f = \frac{1}{2}(3 <\cos^2\theta> -1) \quad (13)$$

$$<\cos^2\theta> = \frac{\int_0^{\pi/2} I(\theta) \cos^2\theta \sin\theta \, d\theta}{\int_0^{\pi/2} I(\theta) \sin\theta \, d\theta} \quad (14)$$

where $\theta$ is the angle between the structural unit vector and the reference direction, and $I(\theta)$ is the anisotropy intensity profile obtained in the $\theta$ range from 0 to $\pi/2$. Therefore, we first determined the peak position in Fig. 6 and used it as a reference to calculate the HOF in the range of 0–$\pi/2$. In the frequency plot produced by ImageJ, the low-frequency pixels are placed in the centre, and the radius corresponds to a frequency in the phase space. The low-frequency signals represent the domains within the original microscopy image that contains pixels with similar intensities, and the intensities of the adjacent pixels do not significantly differ. The bulk of this information originates from the background and overall shape of the image. In contrast, the high-frequency pixels are located away from the origin and close to the periphery of the frequency plot. These pixels form spatial domains that exhibit abrupt changes in the pixel intensity caused by the image edges and noise. Therefore, the circle with a suitable radius was used for low-pass filtering, and the calculated HOF magnitudes exhibited alignment degrees varying from 0 to 0.47 (Fig. 8).

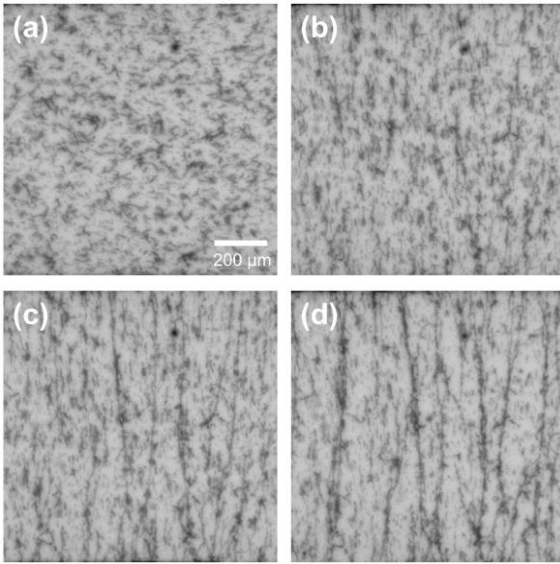

*Fig. 5. Optical images of the studied sample obtained at different stages after applying a 10-kHz electric field*

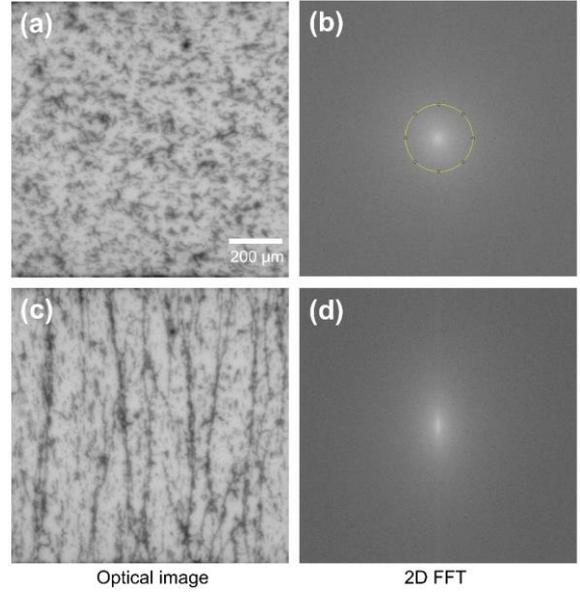

*Fig. 6. Optical images and corresponding 2D FFT intensity profiles.*
*(a) and (c) typical optical images of the studied sample, (b) and (d) the corresponding 2D FFT intensity profiles*

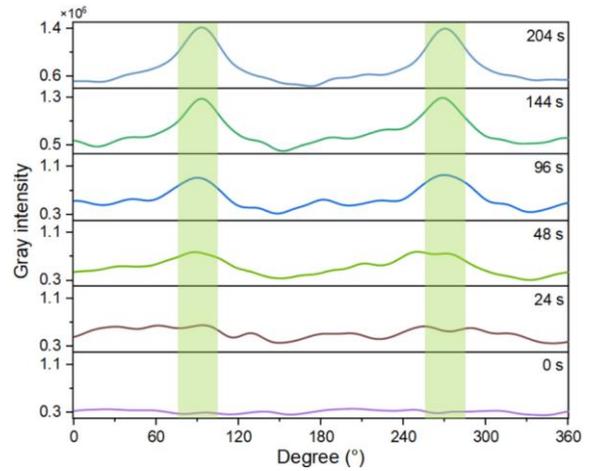

*Fig. 7. Distributions of the spectral intensity along the specified direction obtained at different times*

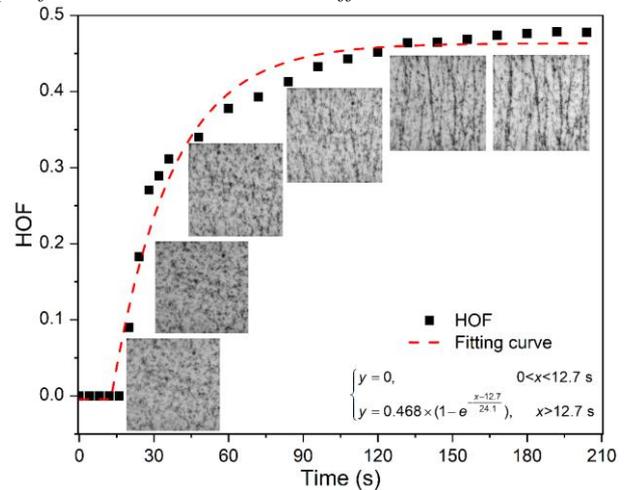

*Fig. 8. HOF changes with time observed after the voltage application*

The results presented in Fig. 6 indicate that after the voltage application, the observed changes in HOF occur in three different stages. At the first stage (corresponding to a time frame of 0–15 s), HOF remains zero; during the second stage, HOF gradually increases (first rapidly and then slowly); finally, at the third stage, the HOF value remains constant at a value of approximately 0.47. This trend can be explained by the optical observation data depicted in Fig. 9. At the first stage, each randomly distributed TiO$_2$ whisker rotates around its respective centroid along the direction of the electric field. These subtle changes are reflected by the high-frequency signals displayed in the 2D FFT frequency plot away from the centre, which are filtered out during spectral intensity calculations. Therefore, the HOF magnitude remains zero. After that, the individual whiskers oriented along the electric field interact with the neighbouring whiskers and connect end-to-end in the direction of the electric field to form short chains. The local short chains also attract each other and connect end-to-end to increase their lengths until they reach both electrodes. As a result, the HOF value gradually increases. When chains bridge the electrodes, both the overall image shape and HOF no longer change. According to the microscopic image, the free particles located near the chains are attracted by them and become deposited on the chain surface to increase the chain strength.

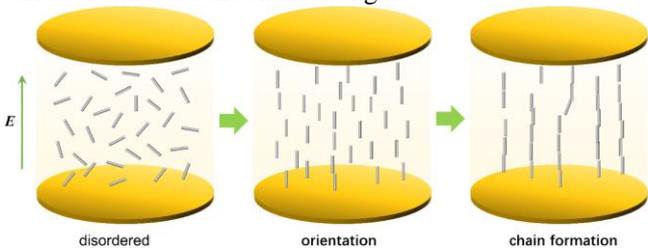

*Fig. 9. Three stages of the self-assembly of rod-like particles under an applied AC electric field*

To model this complex dynamic process, HOF function fitting was performed in the time domain using the following one-phase exponential association equation with a plateau:

$$\text{HOF} = \begin{cases} 0, & 0 < x < 12.7 \text{ s} \\ 0.468 \times \left(1 - e^{\frac{x-12.7}{24.1}}\right), & x \geq 0 \end{cases} \quad (15)$$

### 4.2. Online dielectric permittivity monitoring

This part of the experiment was performed using the device depicted in Fig. 4b. In the first step, we applied voltage to the studied mixture at room temperature and simultaneously monitored the voltage and current signals within 30 min. The real part of the dielectric permittivity was calculated according to (2). After 30 min, the sample was transferred into an oven and heated to a temperature of 125 °C at a rate of 5 °C/min and then cured for 2 h. The voltage was applied throughout the entire process.

Samples with different filler contents were examined at various field strengths. During the voltage application, the output voltage drops slightly, while the current gradually increases. According to (10), the permittivity of the sample increases with time. The real parts of the real-time variable permittivity are plotted in Fig. 10. At t = 0 s (before the electric field application), the higher volume fraction increases the permittivity value (the detection voltage is small and does not affect the filler). This is a well-known phenomenon, although the majority of previous studies on this topic were performed at higher filler volume fractions. Under the applied AC electric field, the permittivity dynamics varies with the volume fraction and field strength. The permittivities of all samples increased with time, and only three samples subjected to a field of 25 V/mm had not reached a constant permittivity within the first 30 min. The obtained results show that the stronger the applied electric field, the faster the permittivity increase because increasing the electric field strength increases the rotational torque and chain force in (3) and (8) according to the second-order power law. In addition, the permittivity increase rate is proportional to the mixture loading for the following two reasons. (I) The particles become closer to each other at a higher volume fraction, which increases the interaction force. (II) The time required for a given particle to reach a neighbouring particle decreases, which promotes the chain growth.

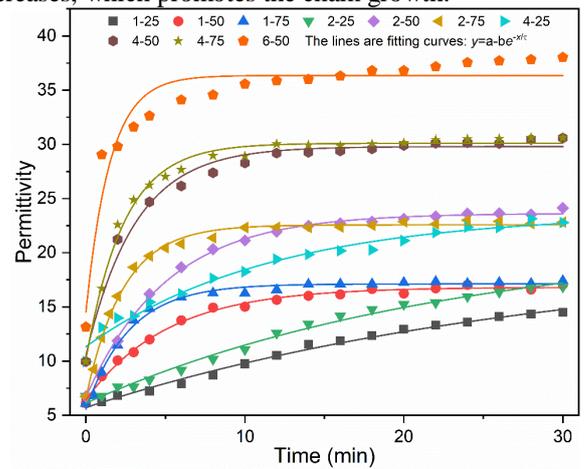

*Fig. 10. Variation of the real part of the sample permittivity with time observed after the voltage application under different experimental conditions. Notation 1-25 represents the filler content of 1 vol.% and applied 10-kHz electric field of 25 V/mm*

The dielectric permittivity spectra of the cured samples are displayed in Fig. 11. They show that at the same filler content, the permittivities of the samples (both their real and imaginary parts) exposed to the electric fields of 50 and 75 V/mm are approximately the same, indicating that their particles have achieved perfect alignment before curing. However, the permittivities of the samples treated with the 25 V/mm electric field were lower than those of the samples exposed to the 50 and 75 V/mm fields, indicating that their particles were not fully aligned before curing. As the filler content increases, the dielectric loss factor increases as well, which should be considered in practical applications.

We also compared the 10-kHz dielectric permittivity of the samples treated with a 50 V/mm electric field at 30 min and after curing, as shown in Fig. 11c. The latter value is slightly lower than the former one due to the stronger polarisation of the liquid matrix. However, both magnitudes are much higher than the values obtained for the cured samples in the absence of an electric field. The permittivity increases almost linearly when the filler content increases from 1 to 6 vol.%. This indicates that the TiO$_2$ whiskers and the matrix are in parallel in the direction of the electric field[26].

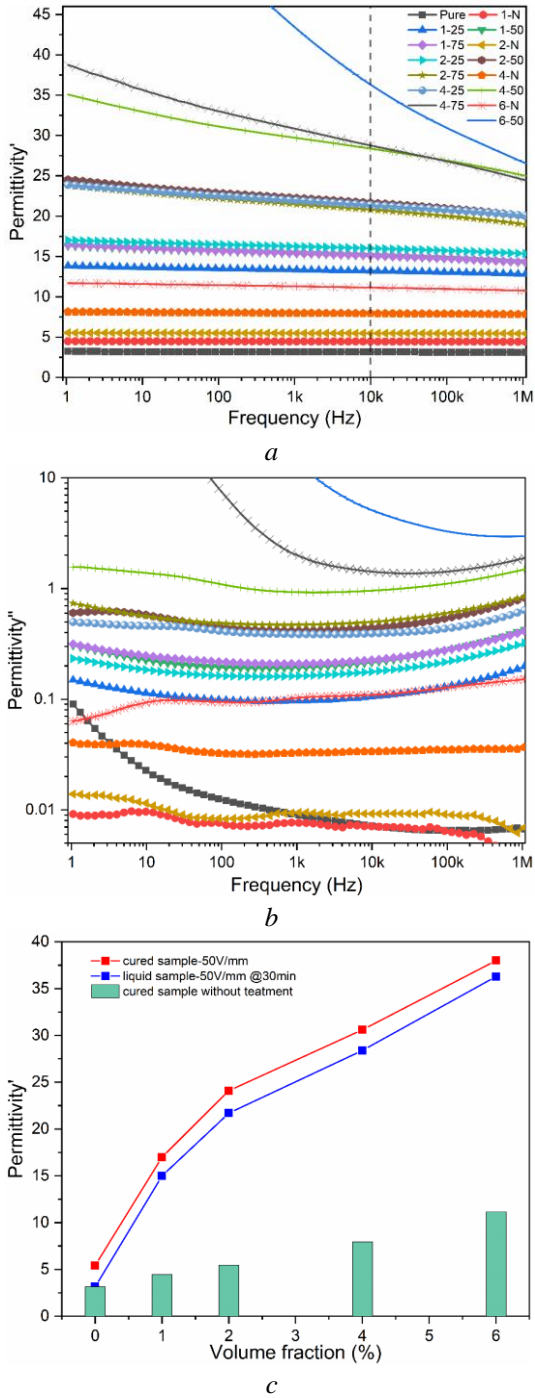

*Fig. 11. Dielectric spectrum test results.*
*(a) real part of the dielectric permittivities of different cured samples,*
*(b) imaginary part of the dielectric permittivities of different cured samples,*
*(c) permittivity values of the samples exposed to different electric fields with a frequency of 10 kHz*

### 4.3. Analytical model for transient permittivity

Optical and permittivity monitoring is carried out in the previous sections. To establish the time-domain model of the transient dielectric permittivity, the experimental data were fitted using the exponential function (16). The obtained fitting parameters are plotted in Fig. 12a.

$$f(t) = a - be^{-\frac{t}{\tau}} \quad (16)$$

Here, $b-a$ and $a$ represent the initial and stable values of the dielectric permittivity under an applied electric field, respectively, and $\tau$ is the time constant reflecting the rate of particle self-assembly that is related to the spacing between the particles and the applied force. Spacing $\lambda$ is proportional to $\phi^{1/3}$, the chain-forming force is proportional to $E^2$, and the attraction force is proportional to $\lambda^{-4}$ for the two isolated parallel dipoles. The increase in the interfacial area increases the mixture viscosity because the surface area is proportional to $\phi^{2/3}$. Using this approach, multiple linear regression analysis was performed for $\ln\tau$, $\ln\phi$, and $\ln E$, as shown in (17) and Fig. 12b. Here, (17) was substituted into (15) to obtain (18), which was utilised to determine parameters a and b at a given volume fraction.

$$\ln\tau = 11.59 - 0.5\ln\phi - 2\ln E \quad (17)$$

$$f(\phi, E, t) = a - be^{-\frac{\phi^{0.5}E^2 t}{107975}} \quad (18)$$

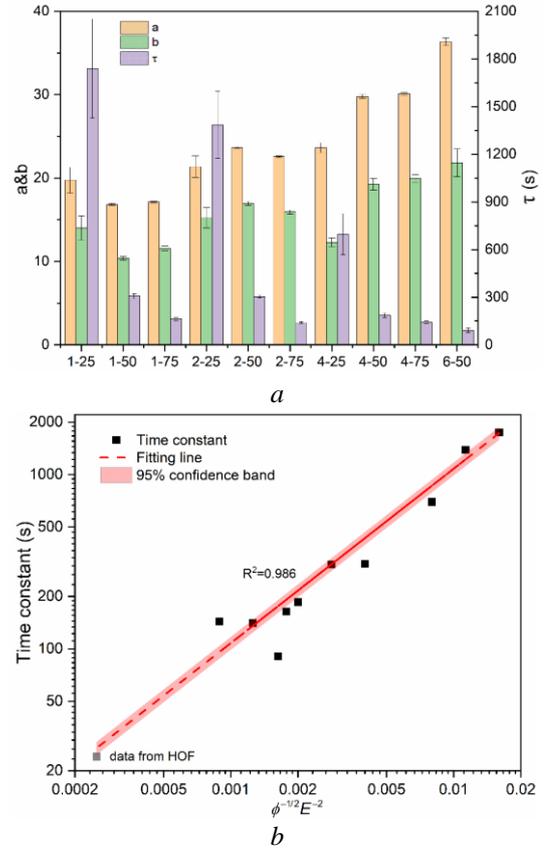

*Fig. 12. Calculated results from Fig. 8 and Fig. 10.*
*(a) exponential fitting parameters of the experimental data presented in Fig. 10,*
*(b) linear regression of the time constant in the proposed transient permittivity model*

By applying this model, we can simulate the spatiotemporal characteristics of dielectric permittivity when an AC voltage is applied to the electrodes attached to an uncured insulating device. Initially, the applied voltage generates a non-uniform electric field, which drives the particle self-assembly process. As a result, the local permittivity of the mixture increases with time. At any given moment, the permittivity has a non-uniform distribution in space. The local permittivity in the region subjected to a stronger electric field is larger than that of the region exposed to a weaker electric field. At the same time, the increased permittivity limits the local electric field strength. Overall, the described transient permittivity model is a complex spatiotemporal model with multi-physical field coupling.

The actual insulation materials often contain large amounts of inorganic reinforcing fillers (such as nano-$SiO_2$ and micro-$Al_2O_3$), which can inhibit the self-assembly of target fillers. When using casting devices, the mixture viscosity may be reduced by warming or other methods to ensure smooth sample degassing and casting despite the presence of reinforcing fillers. For example, the basin-type insulator in GIS is made of epoxy resin with an Al2O3 filler content of 330 phr. Its degassing and casting were performed at a temperature of 130 °C and mixture viscosity of less than 3 Pa·s (DC 184 in this work is 3.5 Pa·s). Therefore, the proposed model is suitable for hybrid systems with reinforcing fillers by modifying the time constant. In addition, these fillers are weaklier polarised and respond to the electric field more slowly than the target fillers; therefore, their self-assembly effect within the operating timeframe is negligible.

Moreover, the dielectrophoretic forces caused by the applied electric field gradient are much weaker than the rotational and chain-forming forces because the majority of the electric fields generated in power systems or electronic components are slightly non-uniform and have small gradients. Therefore, the dielectrophoretic effect during operation can also be neglected.

In summary, the proposed analytical model for the transient dielectric permittivity is suitable for the electric field-assisted preparation of permittivity gradient insulating devices. It considers the spatiotemporal coupling of the permittivity of the uncured composite device subjected to an electric field.

## 5. Conclusion

In summary, a new analytical model for the transient permittivity of uncured $TiO_2$ whisker/LSR composites exposed to an AC electric field is proposed in this paper. We experimentally quantify the self-assembly of target fillers based on optical image and dielectric permittivity monitoring, which avoids the direct processing of complex electrodynamics. Compared with the finite element method (FEM) based on differential equations, this statistical model is simple but efficient, and can be applied to any low-viscosity uncured composites, which may contain multiple fillers. The time constant in this model indicates the self-assembly speed of the target filler and the change rate of the effective permittivity, which is related to the filler content and electric field strength applied. For an uncured composite with given components, the increased filler content and electric field strength can accelerate the transient process. Although the proposed model is based on a uniform AC electric field, it can be used in a non-uniform electric field. When a voltage is applied to an uncured composite insulating device (such as IGBT pouring sealant and basin-type insulators), this model can be used to analyse the spatiotemporal permittivity characteristics and optimise the permittivity gradient for electric field grading.

## 6. Acknowledgments

This work was supported by the National Natural Science Foundation of China (No. 51777107) and Shenzhen fundamental research and discipline layout project (No. JCYJ20180508152044145). We would like to thank Editage (www.editage.cn) for English language editing.

## 7. References

[1] Li, J., Liang, H., Chen, Y., *et al*.: 'Promising functional graded materials for compact gaseous insulated switchgears/pipelines', High Voltage, 2020, **5**, (3), pp. 231–240

[2] Wang, C., Sun, Q., Zhao, L., *et al*.: 'Mechanical and Dielectric Strength of Laminated Epoxy Dielectric Graded Materials', Polymers, 2020, **12**, (3), pp. 622–636

[3] Kurimoto, M., Ozaki, H., Tooru, S.: 'FEM simulation of local field enhancement close to lamination interface of permittivity-graded material', Electronics and communications in Japan, 2018, **101**, (6), pp. 48–57

[4] Muneaki, K., Katsumi, K., Masahiro, H., *et al*.: 'Application of Functionally Graded Material for Reducing Electric Field on Electrode and Spacer Interface', IEEE Trans. Dielectr. Electr. Insul., 2010, **17**, pp. 256–263

[5] Katsumi, K., Muneaki, K., Hideki, S., *et al*.: 'Application of Functionally Graded Material for Solid Insulator in Gaseous Insulation System', IEEE Trans. Dielectr. Electr. Insul., 2006, **13**, pp. 362–372

[6] Diaham, S., Valdez, N. Z., Lévêque, L., *et al*.: 'An original in-situ way to build field grading materials (FGM) with permittivity gradient using electrophoresis', in 2nd IEEE International Conference on Dielectrics (ICD), Budapest, Hungary, 2018, pp. 1–4

[7] Almasi, D., Sadeghi, M., Lau, W.J., *et al*.: 'Functionally graded polymeric materials: A brif review of current fabrication methods and introduction of a novel fabrication method', Mater Sci Eng C Mater Biol Appl, 2016, **64**, pp. 102–107

[8] Naebe, M., Shirvanimoghaddam, K.: 'Functionally graded materials: A review of fabrication and properties', Applied Materials Today, 2016, **5**, pp. 223–245

[9] Li, X.R., Liu, Z., Li, W.D., *et al*.: '3D printing fabrication of conductivity non-uniform insulator for surface flashover mitigation', IEEE Trans. Dielect. Electr. Insul., 2019, **26**, (4), pp. 1172–1180

[10] Shen, Z., Wang, X., Zhang, T., *et al*.: 'In situ electric field driven assembly to construct adaptive graded permittivity BaTiO3/epoxy resin composites for improved insulation performance', Applied Materials Today, 2020, **20**, p. 100647(1–11)

[11] Benhal, P., Chase, J.G., Gaynor, P., *et al*.: 'AC electric field induced dipole-based on-chip 3D cell rotation', Lab on a Chip, 2014, (14), pp. 2717–2727

[12] Hoshino, K., Chen, P., Huang, Y., *et al*.: 'Computational analysis of microfluidic immunomagnetic rare cell separation

from a particulate blood flow', Anal. Chem., 2012, **84**, (10), pp. 4292–4299

[13] Arguin, M., Sirois, F., Therriault, D.: 'Electric field induced alignment of multiwalled carbon nanotubes in polymers and multiscale composites', Advanced Manufacturing: Polymer & Composites Science, 2015, **1**, (1), pp. 16–25

[14] Batra, S., Unsal, E., Cakmak, M.: 'Directed Electric FieldZ-Alignment Kinetics of Anisotropic Nanoparticles for Enhanced Ionic Conductivity', Advanced Functional Materials, 2014, **24**, (48), pp. 7698–7708

[15] Brown, D., Kim, J., Lee, H., *et al.*: 'Electric field guided assembly of one-dimensional nanostructures for high performance sensors', Sensors (Basel), 2012, **12**, (5), pp. 5725–5751

[16] Hoyer, H., Knaapila, M., Helgesen, G., *et al.*: 'Microelectromechanical strain and pressure sensors based on electric field aligned carbon cone and carbon black particles in a silicone elastomer matrix', Journal of Applied Physics, 2012, **112**, (9), pp. 094324(1–8)

[17] Ross, B., Waldeisen, J., Wang, T., *et al.*: 'Strategies for nanoplasmonic core-satellite biomolecular sensors: Theory-based Design', Appl. Phys. Lett., 2009, **95**, (19), pp. 193112(1–3)

[18] Ronald Pethig: 'Review Article—Dielectrophoresis: Status of the theory, technology, and applications', Biomicrofluidics, 2010, **4**, pp. 1–35

[19] Peng, N., Zhang, Q., Li, J., *et al.*: 'Influences of ac electric field on the spatial distribution of carbon nanotubes formed between electrodes', J. Phys. D: Appl. Phys., 2006, **100**, (2), pp. 024309(1–5)

[20] Singh, S., Aryaan, N., Shikder, M., *et al.*: 'A 3D nanoelectrokinetic model for predictive assembly of nanowire arrays using floating electrode dielectrophoresis', Nanotechnology, 2019, **30**, (2), pp. 025301(1-12)

[21] Daniel, J., Ju, L., Yang, J., *et al.*: 'Pearl-Chain Formation of Discontinuous Carbon Fiber under an Electrical Field', Journal of Manufacturing and Materials Processin*g*, 2017, **1**, (2), pp. 1–14

[22] Giner, V., Sancho, M., Lee, R., *et al.*: 'Transverse dipolar chaining in binary suspensions induced by rf fields', J. Phys. D: Appl. Phys., 1999, **32**, (10), pp. 1182–1186

[23] Ayres, C., Jha, B., Meredith, H., *et al.*: 'Measuring fiber alignment in electrospun scaffolds: a user's guide to the 2D fast Fourier transform approach', Journal of Biomaterials Science, Polymer Edition, 2008, **19**, (5), pp. 603–621

[24] Xu, M., Futaba, D., Yumura, M., *et al.*: 'Alignment Control of Carbon Nanotube Forest from Random to Nearly Perfectly Aligned by Utilizing the Crowding Effect', ACS Nano, 2012, **6**, (7), pp. 5837–5844

[25] Gómez, J., Overberghe, N., Tortosa, P., *et al.*: 'Assessment of the parameters influencing the fibre characteristics of electrospun poly(ethyl methacrylate) membranes', European Polymer Journal, 2011, pp. 119–129

[26] Guo, R., Roscow, J., Bowen, C., *et al.*: 'Significantly enhanced permittivity and energy density in dielectric composites with aligned $BaTiO_3$ lamellar structures', J. Mater. Chem. A, 2020, **8**, (6), pp. 3135–3144